# Fault Detection Method for Power Conversion Circuits Using Thermal Image and Convolutional Autoencoder

Noboru Katayama and Rintaro Ishida

*Abstract*—A fault detection method for power conversion circuits using thermal images and a convolutional autoencoder is presented. The autoencoder is trained on thermal images captured from a commercial power module at randomly varied load currents and augmented image2 generated through image processing techniques such as resizing, rotation, perspective transformation, and bright and contrast adjustment. Since the autoencoder is trained to output images identical to input only for normal samples, it reconstructs images similar to normal ones even when the input images containing faults. A small heater is attached to the circuit board to simulate a fault on a power module, and then thermal images were captured from different angles and positions, as well as various load currents to test the trained autoencoder model. The areas under the curve (AUC) were obtained to evaluate the proposed method. The results show the autoencoder model can detect anomalies with 100% accuracy under given conditions. The influence of hyperparameters such as the number of convolutional layers and image augmentation conditions on anomaly detection accuracy was also investigated.

## I. INTRODUCTION

POWER conversion circuits exist at various scales and are used in a wide range of applications, including electronic devices and energy supply systems. While the reliability of these circuits has improved year by year, failures in components such as semiconductor switching devices [1] and electrolytic capacitors remain unavoidable, and are still among the most common failure types [2]. The failure rate is influenced by factors such as initial components defects and temperature conditions. When a power circuit malfunctions, it can cause the entire system to cease operation, making the early detection of anomalies a critical issue. For instance, in photovoltaic systems, approximately 54% of malfunctions occur in the power conditioning system [3], which not only halts the entire system but may also lead to severe accidents, such as fires in the power conditioning system.

Various approaches have been proposed to detect faults in power conversion circuits to date. Sensor-based approaches generally offer fast detection since they can directly sense the fault event as it occurs and are often simple in concept [4]. Thermal images captured with IR cameras are commonly used to detect malfunctions in electronic circuits during both development and maintenance. The temperature of electronic components tends to change when they experience faults. IR cameras are affordable and readily available, offering the advantage of quickly identifying obvious anomalies by simply detecting areas of unusually high or low temperatures [5]. However, since most electronic components naturally generate heat during normal operation, it can be challenging to detect subtle abnormalities. This is especially true for power circuits, where switching devices inherently produce significant heat, and the amount of heat generated can vary depending on the electrical load of the circuit. As a result, detecting subtle anomalies using infrared cameras can sometimes be challenging.

Recently, machine learning approaches have become widely applied across various domains, not only for image and speech recognition and large language models, but also for anomaly detection. Aljameal et al. detected minor leaks in oil or gas pipelines using machine learning-based anomaly detection, and compared five classical machine learning techniques including random forest, support vector machine [6]. Deep learning-based approaches enhance the ability to recognize high-dimensional data such as image data. Tong et al. proposed an object detection algorithm based on convolutional neural networks to extract defect features of composite materials from thermal images [7]. Santos et al. employed deep learning and thermal images to classify 11 types of faults in electric motors using transfer learning and attained an accuracy rate of over 98% [8]. Manno et al. used convolutional neural networks to detect hotspots on photovoltaic cells from images captured by unmanned aerial vehicles [9]. Training neural networks as classifiers requires a large amount of labeled data for training, which means that a pairs of input data (e.g., image data) and corresponding labels (e.g., fault type or mode) are needed.

Unsupervised learning, which does not rely on labeled data for training, has frequently been employed in anomaly detection tasks. In particular, autoencoders have gained increasing attention in deep learning-based unsupervised learning approaches, as they can be trained using only normal data. This characteristic makes them especially effective in scenarios where obtaining a large amount of anomalous data is challenging. Various applications of autoencoders have been proposed in many fields. Haidong et al. used an autoencoder model to diagnose rotating machinery faults in vibration signals [10]. They applied a novel loss function and an optimization technique for hyperparameter tuning to enhance the performance of the model. The autoencoder was used to extract features from the vibration signals, and a softmax layer was employed to classify the type of fault. Chow et al. [11] have reported the application of a convolutional autoencoder to detect defects in concrete structures, and compared their approach with other classical methods. Recently, the autoencoder has been widely applied to medical applications [12], [13], [14], [15], [16], [17], psychology [18], and industrial technology [19].

Several studies have also demonstrated anomaly detection using thermal images acquired by thermal cameras. Fanchiang et al. have presented an online fault monitoring system for cast-resin transformers and proposed an overheating fault diagnosis method based on thermal images [20]. They constructed an autoencoder to detect transformer faults and a classification model to identify the types of fault. Oliveira et al. proposed a method that detects faults using an autoencoder model and thermal images [22]. The maximum temperature in the region of interest were extracted from thermal images, and the spectrum derived from this temperature was calculated. The trained autoencoder reconstructs the fault-free spectrum, which is then compared with the original spectrum to detect anomalies. They also employ a random forest model for comparison; however, the random forest approach has the drawback that anomalous data is required for training. The autoencoder approach is effective in case where limited anomalous data is available.

Identifying the specific areas where faults occur is essential for fault detection in electrical circuits, as numerous components are implemented on a circuit board, and each poses a potential risk of failure. Although the method proposed by Oliveira et al. [22] specifies a region of interest that may contain faults, other components may also exhibit anomalies. For example, in power conversion circuits like power modules, the temperature distribution can exhibit various patterns depending on the load conditions. Therefore, the ability to accommodate such variations is desirable. Furthermore, thermal images may be captured from various angles and positions due to the non-fixed positioning of thermal cameras. Therefore, the ability to detect faults using thermal images acquired under diverse conditions is highly beneficial. Training the autoencoder under various conditions may help meet these requirements. Furthermore, by directly using thermal images for training, it becomes feasible to compare the captured and reconstructed images, thereby facilitating the identification of anomalous regions.

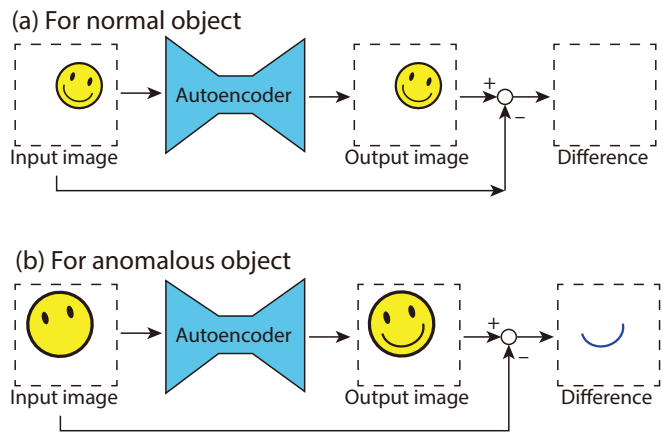

Fig. 1 Process of anomaly detection using an autoencoder.

In this study, a fault detection method for power conversion circuits using a convolutional autoencoder trained on thermal images captured from thermal camera is proposed, and accuracy of the detection is evaluated. A commercially available power module was selected as a sample power conversion circuit, and thermal images of the power module were captured. The autoencoder was trained using thermal images of the power module in a fault-free conndition.

II. METHODOLOGY

The autoencoder is designed to reconstruct its input as accurately as possible; however, this reconstruction ability is effective primarily for inputs that are similar to those seen during training. An autoencoder consists of two main components: an encoder and a decoder. The encoder compresses the input image into a lower-dimensional latent representation, capturing the essential features of the data. The decoder then reconstructs the original image from this compressed representation. During training, the autoencoder learns to minimize the reconstruction error between the input and output images. Because it is trained only on normal data, the autoencoder learns to represent and reproduce only the distribution of normal features, making it sensitive to deviations introduced by anomalous inputs.

Fig. 1 illustrates the process of anomaly detection using an autoencoder trained on images of normal objects. In the case where the input image contains a normal object, as shown in
Fig. 1 (a), which resembles the training data, the output image generated by the autoencoder closely matches the input image. In contrast, when the input image contains an anomalous object as shown in
Fig. 1 (b), the autoencoder, having not encountered such anomalies during training, fails to reconstruct the anomalous regions accurately. Instead, it produces an output that reflects the learned characteristics of normal data. By calculating the difference between the input and output images, it is possible to identify the presence of anomalies and localize their positions within the image. This approach enables effective unsupervised anomaly detection, particularly in settings where anomalous samples are scarce or difficult to label.

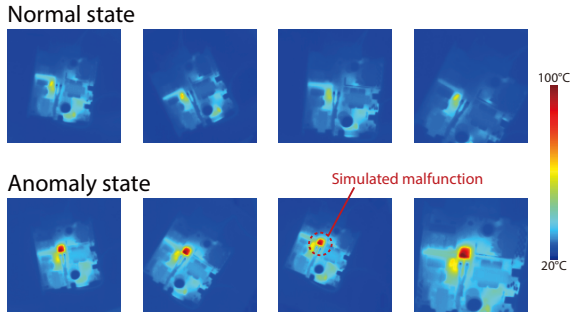

Fig. 6 Example of the acquired thermal images of the power module in normal state and anomaly state by means of the thermal camera.

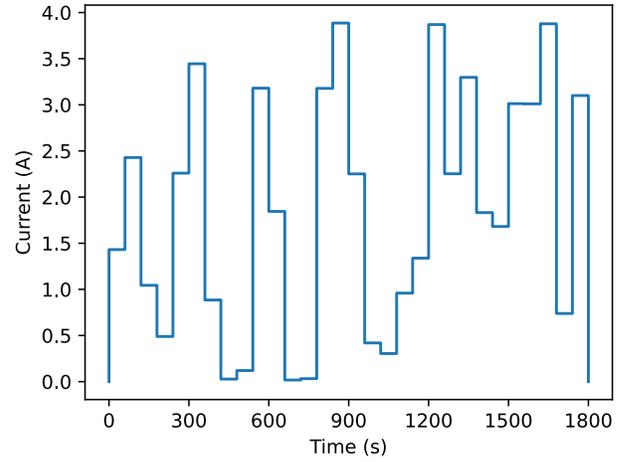

Fig. 4 Current pattern of the electronic load for the training.

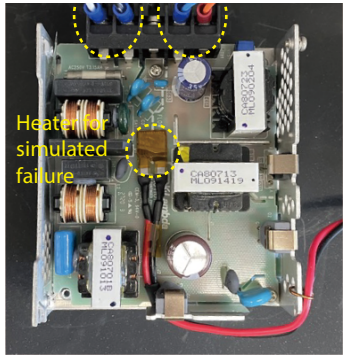

Fig. 3 Photograph of the power module and heater attached to the power module to simulate failure

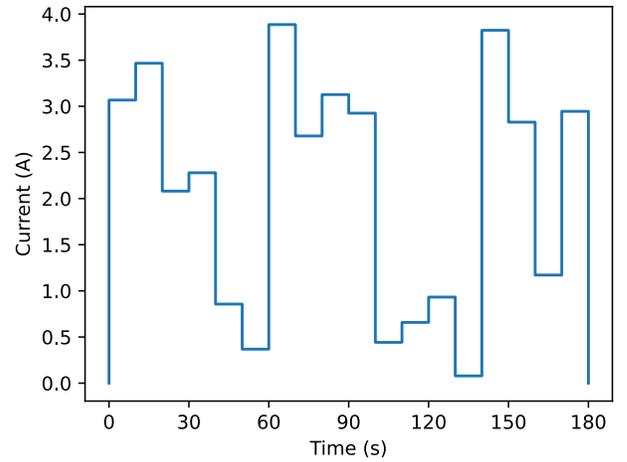

Fig. 5 Current pattern of the electronic load for the testing.

### A. Equipment

Fig. 2 illustrates the experimental setup to capture thermal images of a power module. The thermal camera used in this study was a Lepton FS (FLIR), which has a resolution of 160 x 120 pixels and a frame rate of 8.7 Hz. The horizontal and diagonal field of view are 57° and 71°, respectively. The dynamic range of the captured temperature is -10 to 140 °C and the resolution is 14 bits. A PureThermal 3 breakout board was used to connect the camera to a PC via a USB interface. A power module (RWS100B-24, TDK-Lambda Corp.) used in this study converts AC input voltage ranging from 100 V to 240 V into a DC output with a rated voltage of 24 V. In this setup the module was connected to a AC 100 V power supply as an input and a DC electronic load as an output. The electronic load was operated in constant current mode, with the current limited to 4 A, so that the power module does not exceed the module's rated power of 100 W. A heat insulator and a blackbody board were placed underneath the power module. The blackbody board was prepared by applying blackbody paint to a styrene foam board using a spray method, and was used to minimize thermal reflections captured by the thermal camera, while the heat insulator was intended to reduce heat conduction from the power module to the blackbody board. The power module can be moved with the heat insulator and the blackbody board to adjust the imaging area and angle.

In this experiment, the thermal distribution of the power module is observed while varying the electronic load. The data collected by the thermal camera was processed on the PC, where changes in the heat distribution pattern due to load variations are analyzed. To evaluate the precision of anomaly detection for power modules in this study, a malfunction of the power module was simulated by placing a small heater on the printed circuit board of the power module as shown in Fig. 3. The small heater measures 12 mm x 13 mm and is composed of a heating wire with a resistance of 25 Ω and polyimide film.

### B. Data Acquisition

Thermal images are required for both training and test phases. To cover various load states for the power module, thermal images were captured by changing the current of the electronic load according to a specific pattern. The waveform of the current pattern for training and testing is shown in Fig. 4 and Fig. 5, respectively. For training, the current is updated every 60 seconds to a randomly selected value within the range of 0 to 4 A. The total duration of the training is 1,800 seconds. For testing, the current was varied every 10 seconds, and the total duration was set to 180 seconds. Compared to the training pattern, the test pattern had both a shorter interval between current changes and a shorter overall duration. The longer duration and slower variation in the training current pattern were chosen to allow the temperature of the power module to stabilize following each change, and to enable the model to learn a wide range of thermal distributions. In contrast, the shorter durations in the test pattern reflect the fact that it is not necessary to cover as many patterns during testing.

In practical applications, when capturing images of a power module using a thermal camera, it is expected that the images will be taken from various angles and with different fields of view. This study assumes such conditions. Both the training and test phases use images taken from a range of angles and perspectives. For training data, it is necessary to cover a wide variety of image conditions. However, collecting such diverse data using real images is not realistic. Therefore, this study utilizes image augmentation techniques, which will be described in detail in the following section. For test data, real images must be used to evaluate the accuracy of the proposed method. Since the power module used in this study is movable, it is repositioned together with the heat insulator and the blackbody board. When collecting test data, the current supplied to the electronic load is varied, while at the same time the distance, angle, and position relative to the thermal camera are manually adjusted. An example of the acquired thermal images is shown in Fig. 6. Note that although all the images used in this study are of grayscale, they are mapped to color images for better visualization.

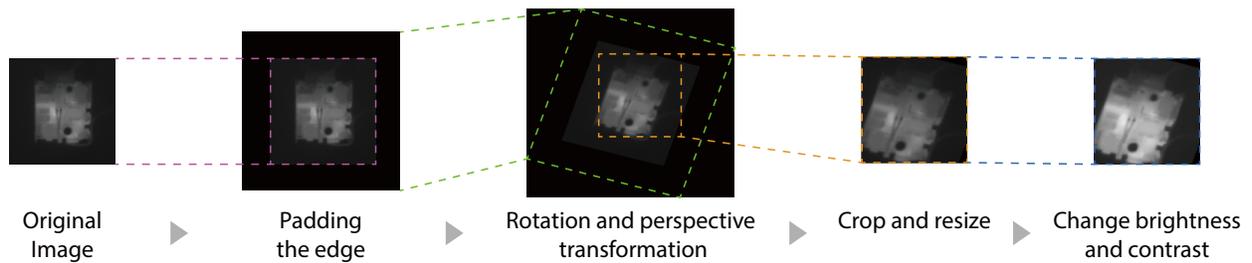

Fig. 7  Process of image augmentation.

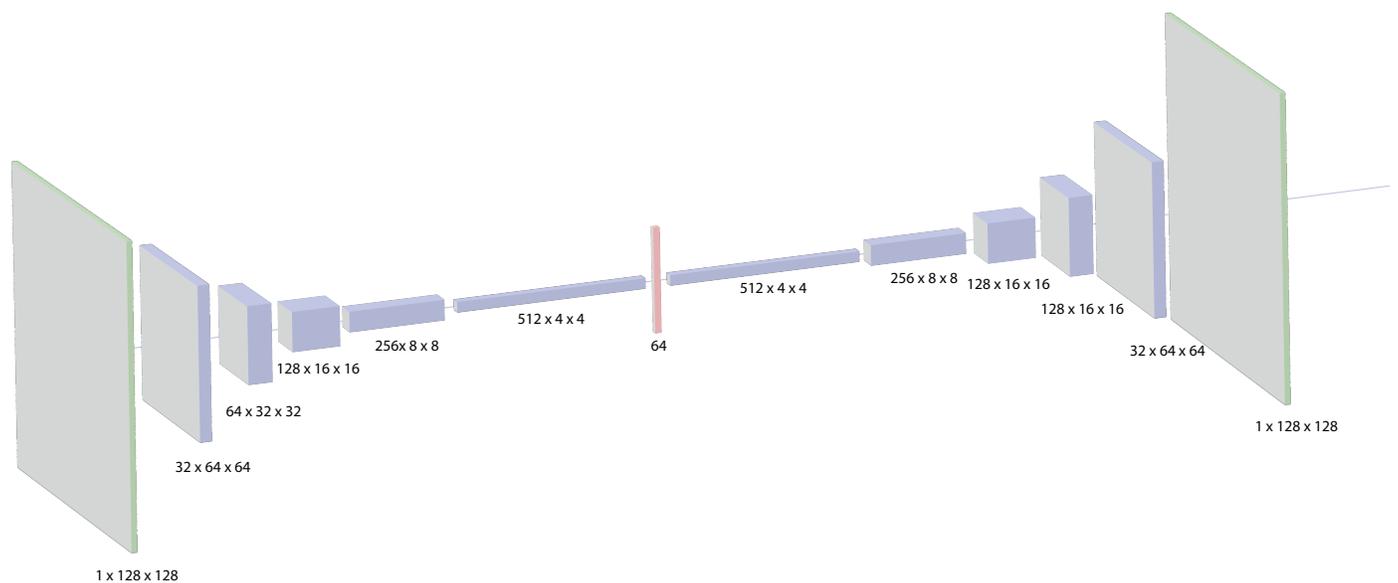

Fig. 8  Structure of the convolutional autoencoder used for fault detection.

*C. Image Augmentation*

In the previous section, we discussed that thermal images of the power module need to be captured from various angles and positions. Therefore, to perform anomaly detection using a data-driven approach with machine learning, images from multiple angles are required. However, in addition to the variations in load types, collecting images taken from various angles and positions is extremely labor-intensive and time-consuming. To address this issue, this study employs data augmentation techniques as shown in Fig. 7.

From 600 thermal images, we generate 10,000 images using image processing. First, an original image is randomly chosen and padded on all edges to resize it to 150% of its original size. Next, the resized image is rotated within a range of -45° to 45°, undergoes a perspective transformation with a scale range of 0.0 to 0.1, and is cropped to a size ranging from 44% to 100% of its original size. Finally, the brightness and contrast of the transformed image are randomly adjusted within ±10%, and the image is resized to 128 x 128. The generated images were split, with 90% used for training and 10% for validation.

*D. Convolutional Autoencoder*

In recent years, autoencoders have been widely used for image compression and reconstruction tasks, enabling efficient feature extraction and dimensionality reduction. In this study, we propose a convolutional autoencoder designed to reconstruct grayscale images of size 128×128 pixels while preserving structural fidelity.

The proposed model consists of an encoder and a decoder, both built using convolutional layers with nonlinear activation functions. The structure of the autoencoder is shown in Fig. 8. The encoder comprises five convolutional layers, each using a kernel size of 3×3 with a stride of 2, progressively reducing the spatial dimensions while increasing the number of feature maps. The activation function ReLU is applied after each convolutional layer to introduce nonlinearity. The final feature maps, sized 4×4×512, are flattened and passed through a fully connected layer to generate a 64-dimensional latent representation. This bottleneck representation serves as a compressed form of the input image, capturing its essential structural features.

The decoder is structured symmetrically to the encoder, utilizing five transposed convolutional layers to progressively upsample the latent representation back to the original spatial resolution. Similar to the encoder, ReLU activations are applied at each layer, except for the final output layer, which uses the sigmoid function to ensure that the pixel values remain within the range of 0 to 1. The reconstruction process aims to minimize the difference between the input and output images while preserving the perceptual quality.

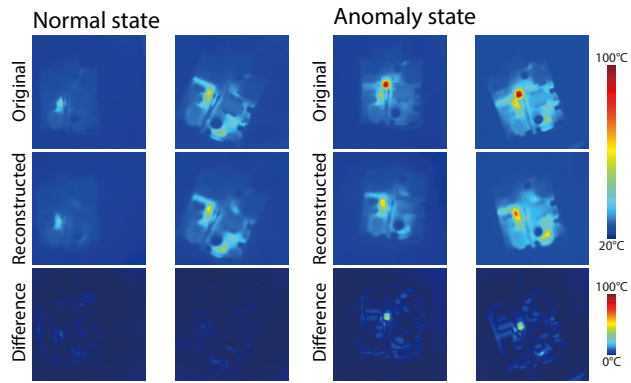

Fig. 9 Original thermal images, reconstructed images and the difference images for normal state and anomaly state of the power module.

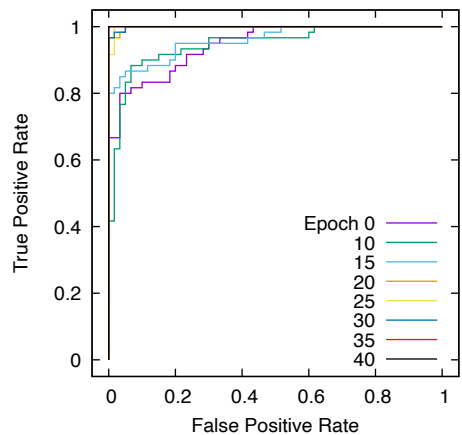

Fig. 10 ROC curves for each epoch.

To train the autoencoder model, the training images were fed at a batch size of 32, which is the number of the images to input at the same time. The number of training epochs was set to 100. The Adam (Adaptive Moment Estimation) optimizer, one of the most widely used optimization algorithms, was employed for parameter updates. Unlike conventional approaches that utilize mean squared error loss, the proposed method adopts the multi-scale structural similarity (MS-SSIM) loss function, which aligns better with human visual perception by considering local contrast, brightness, and structural integrity. To enhance the quality of reconstructed images, the loss function is defined as one minus the MS-SSIM value, encouraging the model to optimize for perceptual similarity rather than pixel-wise differences. This approach effectively reduces artifacts and enhances the preservation of fine details. By leveraging MS-

SSIM as the optimization criterion, the proposed autoencoder is expected to achieve superior performance in image reconstruction compared to conventional MSE-based methods.

## III. RESULTS AND DISCUSSION

### A. Fault detection

The example of the original thermal images, the reconstructed images obtained from the originals using of the autoencoder, and the difference images between the original and reconstructed images for normal state and anomaly state of the power module are shown in Fig. 9. The temperature on the power module in the normal state is approximately between 20°C and 60°C while it reaches up to 100°C in the anomaly state. When the power module is in a normal state, the reconstructed images are almost identical to the original images although the details appear slightly blurred. In the difference image, only areas with changes, such as edges, are faintly visible. Compared with the difference images for the normal state, the difference images for the anomaly state show regions of higher intensity corresponding to the location of the heater used to simulate a fault.

Fig. 10 shows ROC (Receiver Operating Characteristic) curves for each epoch. The ROC curve is a graph used to evaluate the performance of an anomaly detection model. The x-axis and y-axis represent false positive rate and true positive rate, respectively, and the curve is generated by varying the decision threshold to classify the data to normal or anomaly using sampled test data. The area under the ROC curve (AUC) quantifies the performance of the anomaly detection. A higher AUC indicates better model performance: the detection is considered perfect if the AUC is 1.0, and equivalent to random guessing if the AUC is 0.5. As the number of epochs increases and the model training progresses, the ROC curves rise accordingly. By epoch 40, the curve becomes nearly a square, indicating that the AUC approaches 1.0, which indicates that the model can perfectly detect anomalies in the given test data.

Although the current of the heater, which is positioned on the power module to simulate the fault, was set to 0.15 A as a standard condition, the intensity of faults differs in general. To investigate the sensitivity of the proposed method, the heater current was changed. Fig. 11 shows the original, reconstructed, and difference images for different heater currents of 0.15, 0.08, 0.07 and 0.02 A. The hot spot colored in red shown in original images at the heater current of 0.15 A becomes dimmer with the current decreased. While the difference image for the heater current of 0.15 A apparently highlights the fault position, the highlighted positions become less apparent at lower current. Fig. 12 and Fig. 13 show ROC curves and AUC values for the different heater currents, respectively. The model performance degrades as the heater current decreases, suggesting a possible threshold around 0.07 to 0.08 A. The AUC approaches approximately 0.5 when the heat current decreases to 0.02 A

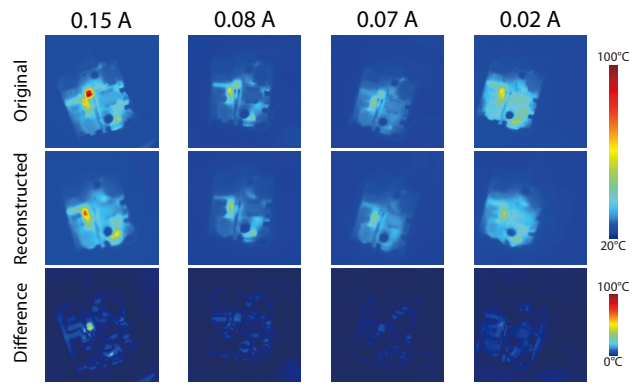

Fig. 11 Original thermal images, reconstructed images and the difference images for anomaly state of the power module at heater currents of 0.15, 0.08, 0.07 and 0.02 A.

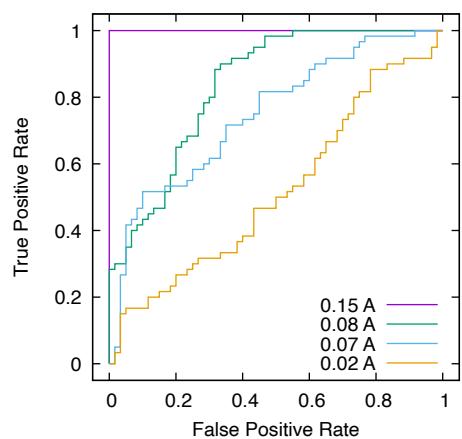

Fig. 12 ROC curves at heater currents of 0.15, 0.08, 0.07 and 0.02 A.

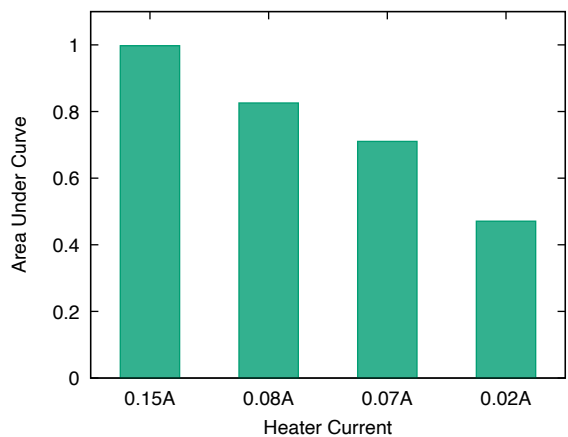

Fig. 13 AUC values at heater currents of 0.15, 0.08, 0.07 and 0.02 A.

### B. Hyperparameter tuning

Hyperparameters are settings chosen prior to training that control how the training process behaves, and it is important to investigate the relationship between hyperparameters and the performance of anomaly detection. In this study, we tuned the hyperparameters to achieve better performance. This study focuses on two representative hyperparameters: the number of convolutional layers and the dimension of the latent space. The number of convolutional layers influences the complexity of the image features that the neural network can represent. Although increasing the number of layers can enhance the reproducibility of the output images, it also leads to greater computational cost and memory usage during training. The dimension of the latent space is related to the level of abstraction achieved by the model, affecting its ability to capture essential features.

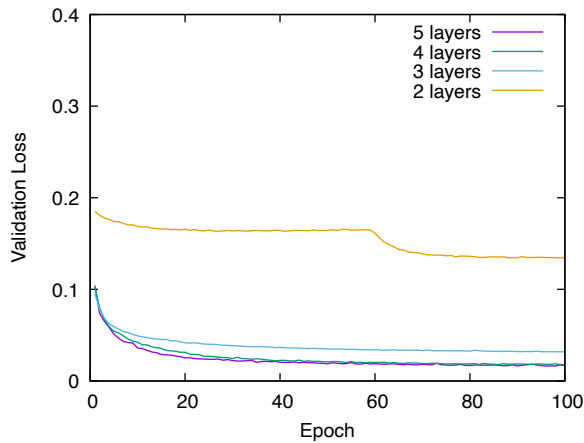

Fig. 14 Validation loss curves over 100 training epochs for models with different number of convolutional layers.

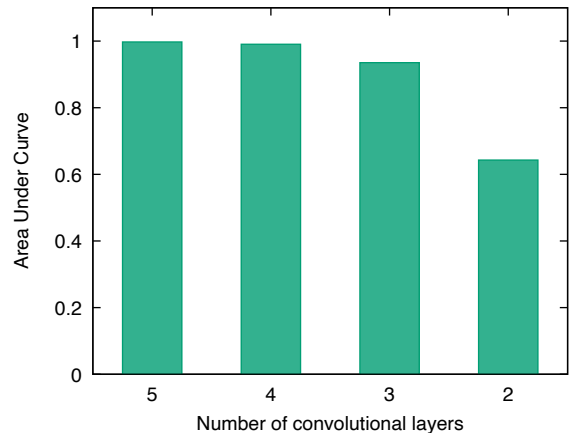

Fig. 16 AUC for models with different numbers of convolutional layers.

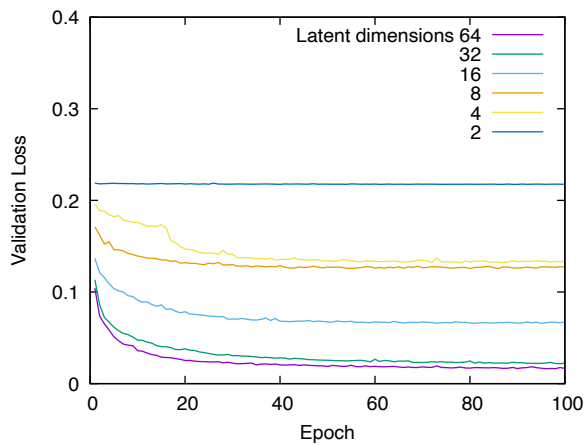

Fig. 15 Validation loss curves over 100 training epochs for models with different number of latent dimensions.

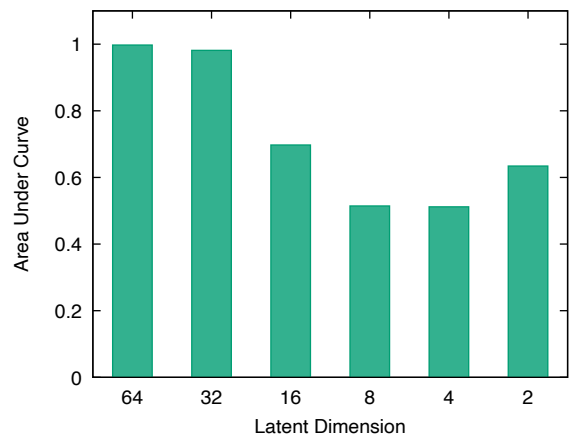

Fig. 17 AUC for models with different latent dimensions.

Fig. 14 and Fig. 15 show validation loss curves over 100 training epochs for models with varying numbers of convolutional layers and latent dimensions, respectively. The validation losses were calculated by applying the loss function to the difference between the model's output and the input using validation images, which were not used for training. This loss reflects how accurately the model reconstructs the input images. Under the condition of this study, autoencoders with a larger number of convolutional layers tend to have lower validation losses. Models with more than four number layers achieved similar validation loss, suggesting that increasing the number of layers beyond four does not significantly improve performance. The validation curves also indicates that 100 training epochs are sufficient, as the loss converge by the end of training. In addition, the latent dimensions also affect the validation loss. The results suggest that a latent dimensions around 32 is sufficient to achieve good reconstruction performance.

To evaluate the performance as an anomaly detection model, the AUC values of the model with different convolutional layers and latent dimensions were calculated and the results are shown in Fig. 16 and Fig. 17. The results are largely consistent with the trends observed in the validation loss. When the AUC approaches 0.5, the model's performance is nearly equivalent to random guessing, indicating that it fails to detect anomalies effectively. In particular, the model with only two convolutional layers and that with a latent dimension of 8 show low AUC values, suggesting that anomaly detection is no longer feasible under these configurations.

### C. Image augmentation

In this study, images are generated from 600 original images to enhance the variation in input image data. As explained in the previous section, original images are randomly selected, followed by resizing, rotation, perspective transformation, and adjustment of brightness and contrast. The number of augmented images and the types of image processing applied can influence the performance of anomaly detection.

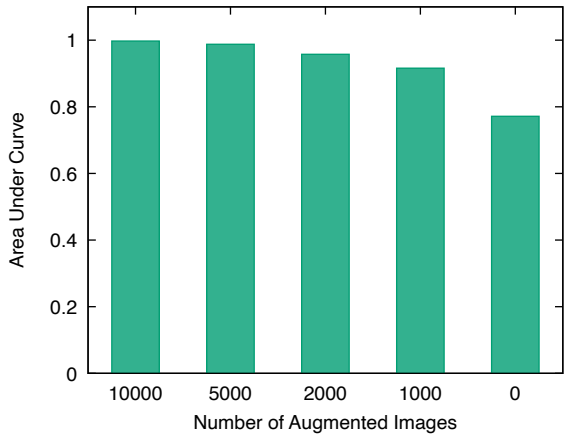

Fig. 18  AUC for models with different number of augmented images.

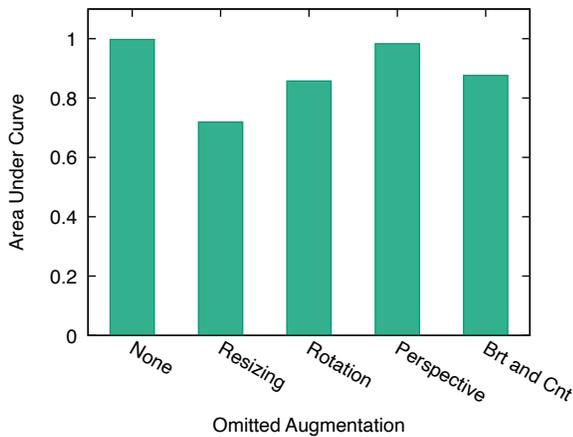

Fig. 19  AUC for models with individual augmentation process omitted.

To investigate this effect, the AUC was evaluated by varying the number of augmented images (10,000, 5,000, 2,000, 1,000 and 0) and by omitting individual augmentation processes, and the results are shown in Fig. 18 and Fig. 19, respectively. Note that the number of augmented images includes the number of original images. A larger number of augmented images leads to better anomaly detection performance. When the number of augmented images is 5,000, the performance appears sufficient; however, the AUC is 0.9877, indicated slight imperfection in anomaly detection. The optimal number likely lies between 10,000 and 5,000. Nevertheless, generating augmented images at this scale does not impose a significant computational burden, so it is recommended to generate as many augmented images as possible.

Regarding the impact of each omitted process on AUC, omitting the process of resizing causes the greatest decrease, followed by rotation, brightness and contrast, and perspective transformation. Therefore, the contribution of each augmentation process can be considered higher in this order. Since thermal images were captured from various angles and positions, it is reasonable that resizing and rotation are important to generate augmented images. The reason adjustment of brightness and contrast had an effect may be that it compensates for slight variations in the overall temperature of the power module as captured in the images. Adjustment of brightness and contrast as an image augmentation technique can be effective in situations where only a limited number of images are available to capture temperature variations. However, perspective transformation had a limited impact on model performance. This may be attributed to the minimal distortion present in the captured image because the thermal camera was fixed, and the power module was kept nearly level throughout capturing. Nevertheless, such augmentation might become necessary if the camera angle were to changes.

## IV. Conclusion

A fault detection method using thermal images and convolutional autoencoder has been proposed and demonstrated using a power module. A convolutional autoencoder is used for an anomaly detection model, bringing a merit that only thermal images of the power module in normal state. During capturing thermal images, the load current was changed randomly to cover wide range of operating conditions. Data augmentation techniques were employed to enable anomaly detection even when the angle or position of the thermal images varied. Anomalies were simulated by attaching a small heater, and under the given conditions, anomaly detection was achieved with nearly 100% accuracy. The effects of hyperparameters such as the number of convolutional layers and the latent dimension, as well as the conditions for image augmentation were also investigated. The results indicate that setting these parameters appropriately is critical for achieving high detection accuracy.

## References

[1] C. Kulkarni, G. Biswas, X. Koutsoukos, K. Goebel, J. Celaya, and M. Field, "PHYSICS OF FAILURE MODELS FOR CAPACITOR DEGRADATION IN DC-DC CONVERTERS," in *The Maintenance & Reliability Conference, MARCON 2010*,

[2] L. P. A. Nathan, R. R. Hemamalini, R. J. R. Jeremiah, and P. Partheeban, "Review of condition monitoring methods for capacitors used in power converters," *Microelectronics Reliability*, vol. 145, p. 115003, Jun. 2023, doi: 10.1016/j.microrel.2023.115003.

[3] P. Kut, K. Pietrucha-Urbanik, and P. Kurek, "Comprehensive Analysis of Failures in Photovoltaic Installations—A Survey-Based Study," *Energies*, vol. 17, no. 23, p. 5986, Nov. 2024, doi: 10.3390/en17235986.

[4] S. S. Khan and H. Wen, "A Comprehensive Review of Fault Diagnosis and Tolerant Control in DC-DC Converters for DC Microgrids," *IEEE Access*, vol. 9, pp. 80100–80127, 2021, doi: 10.1109/ACCESS.2021.3083721.

[5] A. A. Sarawade and N. N. Charniya, "Detection of Faulty Integrated Circuits in PCB with Thermal Image Processing," in *2019 International Conference on Nascent Technologies in Engineering (ICNTE)*, Navi Mumbai, India: IEEE, Jan. 2019, pp. 1–6. doi: 10.1109/ICNTE44896.2019.8946061.

[6] S. S. Aljameel *et al.*, "An Anomaly Detection Model for Oil and Gas Pipelines Using Machine Learning," *Computation*, vol. 10, no. 8, p. 138, Aug. 2022, doi: 10.3390/computation10080138.

[7] Z. Tong, L. Cheng, S. Xie, and M. Kersemans, "A flexible deep learning framework for thermographic inspection of composites," *NDT & E International*, vol. 139, p. 102926, Oct. 2023, doi: 10.1016/j.ndteint.2023.102926.

[8] L. F. Dahmer Dos Santos, J. L. D. S. Canuto, R. C. Thom De Souza, and L. B. R. Aylon, "Thermographic image-based diagnosis of failures in electrical motors using deep transfer learning," *Engineering Applications of Artificial Intelligence*, vol. 126, p. 107106, Nov. 2023, doi: 10.1016/j.engappai.2023.107106.

[9] D. Manno, G. Cipriani, G. Ciulla, V. Di Dio, S. Guarino, and V. Lo Brano, "Deep learning strategies for automatic fault diagnosis in photovoltaic systems by thermographic images," *Energy Conversion and Management*, vol. 241, p. 114315, Aug. 2021, doi: 10.1016/j.enconman.2021.114315.

[10] H. Shao, H. Jiang, H. Zhao, and F. Wang, "A novel deep autoencoder feature learning method for rotating machinery fault diagnosis," *Mechanical Systems and Signal Processing*, vol. 95, pp. 187–204, Oct. 2017, doi: 10.1016/j.ymssp.2017.03.034.

[11] J. K. Chow, Z. Su, J. Wu, P. S. Tan, X. Mao, and Y. H. Wang, "Anomaly detection of defects on concrete structures with the convolutional autoencoder," *Advanced Engineering Informatics*, vol. 45, p. 101105, Aug. 2020, doi: 10.1016/j.aei.2020.101105.

[12] S. Chatterjee *et al.*, "StRegA: Unsupervised anomaly detection in brain MRIs using a compact context-encoding variational autoencoder," *Comput Biol Med*, vol. 149, p. 106093, Oct. 2022, doi: 10.1016/j.compbiomed.2022.106093.

[13] D. Kumar, C. Verma, Z. Illes, A. Mittal, B. Bakariya, and S. B. Goyal, "Anomaly Detection in Chest X-Ray Images using Variational Autoencoder," in *2023 6th International Conference on Contemporary Computing and Informatics (IC3I)*, Gautam Buddha Nagar, India: IEEE, Sep. 2023, pp. 216–221. doi: 10.1109/IC3I59117.2023.10397595.

[14] N. Jahan and Md. A. M. Hasan, "Autoencoder-based Unsupervised Anomaly Detection for Covid-19 Screening on Chest X-Ray Images," in *2022 19th International Conference on Electrical Engineering, Computing Science and Automatic Control (CCE)*, Mexico City, Mexico: IEEE, Nov. 2022, pp. 1–6. doi: 10.1109/CCE56709.2022.9975962.

[15] S. Chouhan and B. Rudra, "Autoencoder-Based Anomaly Detection in ECG Image Time Series Data: A Comparative Evaluation of Three Different Architectures," in *2023 14th International Conference on Computing Communication and Networking Technologies (ICCCNT)*, Delhi, India: IEEE, Jul. 2023, pp. 1–6. doi: 10.1109/ICCCNT56998.2023.10306672.

[16] R. Siddalingappa and S. Kanagaraj, "Anomaly Detection on Medical Images using Autoencoder and Convolutional Neural Network," *IJACSA*, vol. 12, no. 7, 2021, doi: 10.14569/IJACSA.2021.0120717.

[17] X. Zhou *et al.*, "Spatial-contextual variational autoencoder with attention correction for anomaly detection in retinal OCT images," *Comput Biol Med*, vol. 152, p. 106328, Jan. 2023, doi: 10.1016/j.compbiomed.2022.106328.

[18] A. Masaki, K. Nagumo, B. Lamsal, K. Oiwa, and A. Nozawa, "Anomaly detection in facial skin temperature using variational autoencoder," *Artif Life Robotics*, vol. 26, no. 1, pp. 122–128, Feb. 2021, doi: 10.1007/s10015-020-00634-2.

[19] M. Ke, C. Lin, and Q. Huang, "Anomaly detection of Logo images in the mobile phone using convolutional autoencoder," in *2017 4th International Conference on Systems and Informatics (ICSAI)*, Hangzhou: IEEE, Nov. 2017, pp. 1163–1168. doi: 10.1109/ICSAI.2017.8248461.

[20] K.-H. Fanchiang, Y.-C. Huang, and C.-C. Kuo, "Power Electric Transformer Fault Diagnosis Based on Infrared Thermal Images Using Wasserstein Generative Adversarial Networks and Deep Learning Classifier," *Electronics*, vol. 10, no. 10, p. 1161, May 2021, doi: 10.3390/electronics10101161.

[21] Y. Zhang, F. Yu, X. Wang, Q. Zhou, J. Liu, and M. Liu, "Direct operation of Ag-based anode solid oxide fuel cells on propane," *Journal of Power Sources*, vol. 366, pp. 56–64, Oct. 2017, doi: 10.1016/j.jpowsour.2017.08.111.

[22] J. P. Oliveira, C. J. A. Bastos Filho, and S. C. Oliveira, "Non-intrusive Embedded Systems Anomaly Detection using Thermography and Machine Learning," in *Anais do 15. Congresso Brasileiro de Inteligência Computacional*, SBIC, Jan. 2021, pp. 1–8. doi: 10.21528/CBIC2021-20.

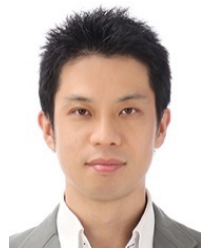

**Noboru Katayama** (Member) received the B.S., M.S., and Ph.D. degrees in engineering from Tokyo University of Science, Japan, in 2006, 2008, and 2011, respectively. He is currently an Associate Professor in the Department of Electrical Engineering, Faculty of Science and Technology, Tokyo University of Science from 2020. His research interests include hydrogen energy, energy device diagnosis, and energy management. Dr. Katayama is a Member of the Institute of Electrical Engineers of Japan, and IEEE

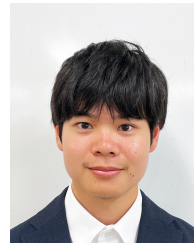

**Rintaro Ishida** (Non-member) received the B.S. degrees in engineering from Tokyo University of Science, Japan, in 2025. He is currently a graduate student in the Department of Electrical Engineering, Faculty of Science and Technology, Tokyo University of Science from 2025. His research interests include fault detection using machine learning.